\documentclass[fleqn,twoside]{article}
\usepackage{espcrc2}

\usepackage{graphicx}
\usepackage[figuresright]{rotating}

\newcommand{\AmS}{{\protect\the\textfont2
  A\kern-.1667em\lower.5ex\hbox{M}\kern-.125emS}}

\title{IceTop Status in 2004
      }

\author{Todor Stanev\address[Bartol]{Bartol Research Institute,
 University of Delaware, Newark, DE 19716, U.S.A.}
        \thanks{The IceCube construction project is supported by the
	Office of Polar Programs of the U.S. National Science Foundation
	with grant OPP-0236449. Please see the web page above for a 
	full list of the funding agencies.}
 for the IceCube Collaboration \address{see {\em http://icecube.wisc.edu}
        for a full list of authors}
	and
        Ralf Ulrich\address{Forschungszentrum Karlsruhe,
            Institut fuer Kernphysik, Postfach 3640, 76021 Karlsruhe, Germany}
        }
       
\begin{document}

\begin{abstract}
IceTop is the surface component of IceCube.  Goals, plans and status of
IceTop in 2004 are reported.
\vspace{1pc}
\end{abstract}

\maketitle

\section{INTRODUCTION}

 IceTop is the surface air shower array above the IceCube~\cite{NAR04}
 neutrino detector.
 Its primary purpose is to complement the detection of high energy
 astrophysical neutrinos and support IceCube by identifying 
 background events. To do that IceTop will~\cite{TKG03}:\\
 1) Tag a fraction of the small showers on the surface 
 that are associated with the main atmospheric
 muon background in IceCube.  Among the identified events will
 be a sample in which two independent
 single muons occur within the reconstruction time interval of
 IceCube at different locations.  Study of such events,
 which have the potential of being
 misreconstructed as  horizontal or upward going muons or showers,
 will be particularly valuable.\\
 2) Veto events with large energy deposition inside IceCube
 when their associated air showers are detected at the surface. 
 IceTop will be able
 to cover a large fraction of the upper hemisphere at energies
 approaching 10$^{17}$ eV because there is no need to 
 reconstruct fully the surface shower to veto an underground event.\\
 3) Provide an independent measure of angular resolution 
 and pointing of IceCube by use of independently reconstructed
 showers detected in coincidence with IceCube.
 This task has been already tested with SPASE2/AMANDA
 coincidences~\cite{NIM04}.   Analysis of coincident events can
 also provide independent information on ice quality.\\

In addition, IceTop together with IceCube constitutes a
novel, three-dimensional air shower array with an aperture approaching
1~km$^2$-sr.  As such it will be able to measure the primary
cosmic-ray spectrum and composition from below the knee 
up to 10$^{18}$ eV.  The high elevation of the South Pole, equivalent
to a vertical depth of 700~g/cm$^2$, has the advantage that
the electromagnetic shower component is observed relatively near
shower maximum, thus minimizing the effects of fluctuations
in the relation between observed size and total energy.
Sensitivity to primary mass comes from measurement of the ratio
of the muon bundle in the deep detector to the shower size at
the surface.  In the process of analyzing SPASE2/AMANDA 
coincident events~\cite{APP04}, Monte Carlo studies showed
 that the number of muons passing through the deep detector
 is well-correlated with the total amount of Cherenkov light observed,
 after accounting for the location of the trajectory of the
 shower core (muon bundle).
 
\section{TESTING AT SOUTH POLE}

 After preliminary tests in 2000 and 2001 with frozen ice-Cherenkov
 tanks instrumented with AMANDA analog optical modules, two 
 test tanks, each containing two early
 versions of the IceCube digital optical modules (DOMs) were deployed
 at the end of 2003.  The two tanks have a surface area of 3~m$^2$
 and ice depths of 0.9 and 1.0 m respectively and are insulated with
 polyurethane foam.
Their rates and the collected
 waveforms demonstrate a good ice quality inside the tanks. 
 
\begin{figure}[thb]
\vspace*{-20pt}
\centerline{\includegraphics[width=70truemm]{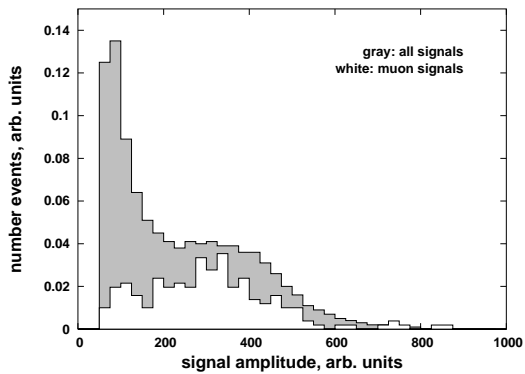}}
\vspace*{-20pt}
\caption{Amplitude distribution of the waveforms of all signals (shaded)
and the muon signals.
\label{serap}
}
\vspace*{-20pt}
\end{figure}

 At the time of deployment the DAQ system was in its early stages and
 the DOM readout was done with a preliminary code with slow
 transmission speed and large dead time. We were nevertheless able to 
 communicate with the DOMs, control their voltage, monitor their temperatures and
 collect waveforms. A fraction of these waveforms was taken with 
 a muon telescope consisting of two 0.2 m$^2$ scintillator counters
 positioned vertically  90 cm apart from each other. The
 identification of the muon signals was made off-line using the GPS
 times of the muon telescope and the DOMs in the tank.
 Fig.~\ref{serap} shows the amplitude distributions for all waveforms
 and for the muon telescope triggers.
The amplitude and charge spectra of tank hits with its characteristic
muon peak (as shown in Fig.~\ref{serap}) 
will be used to monitor and calibrate the IceTop detectors.

 Another important test was the measurement of the temperature of the
 main board of the DOMs during the Antarctic winter. Although the
 DOMs were not always powered there was no failure of the system 
 even when the air temperature dropped suddenly below -70$^\circ$C.  
 As a consequence
 of its thermal mass combined with the tank insulation, the ice
 temperature varied smoothly during the austral winter,
 reaching a low of -55$^\circ$C in August 2004 and increasing gradually
 since then.
The DOM main board typically runs 10$^\circ$C higher than the ice temperature
when the PMT is on.
  
\section{IceTop DESIGN}
 
 IceTop will consist of 80 stations arranged in a triangular grid with 
 average distance between the stations of 125 m. Each station is in the
 vicinity of an IceCube hole, and will share through a junction box
 power supply and communication cables with the under-ice
 detectors. SPASE 2 will be inside the IceTop perimeter and
 will be used in coincidence for at least the initial 3 years of the
 IceCube/IceTop deployment period.

 Each IceTop station will consist of two tanks filled with frozen
 water and viewed by standard IceCube DOMs. A DOM
 consists of 10 inch Hammamatsu R-7081PMT  and 
 boards on which electronics and testing equipment is
 mounted~\cite{spiering}.
 Some of the main board firmware will be modified to suit the
 air shower detection. The two PMT in the same tank will be tuned
 to different gains to enlarge the dynamic range of the 
 tank to about 10$^6$.

 Tanks are made of high-density polyethylene and have radius of 1 m.
 The two DOMs are positioned about 72 cm inward from
 the walls of the tank. The tanks are then filled 
 with 90 cm of water. The walls of the tanks are
 covered with diffusely reflecting (e.g. {\em tyvek}) liners, while 
 the inside tops of
 the tanks are not reflective.

 The ice quality in the tank has to be fairly good so the
 sensitivity to particle signals is uniform throughout the tank.
 A top-down freezing method was developed and tested in Delaware
 and at South Pole. To ensure clear, bubble-free  
 ice the water below the freezing front is degassed with a small
vacuum and pump system inside the tank operated by a
freeze-control unit mounted on the side of the tank that
also controls the ejection of the water produced by
the expansion of the ice.
 The freezing control process is automatically monitored.
 
\section{DATA ACQUISITION}

 The signal rate at the surface is much higher than in ice. 
To manage the high data rate we distinguish between
single particle hits and potential air shower signals
by use of local coincidence between the two tanks at
each station.
A single tank hit is assumed to be 
 a single particle: GeV $\mu $, $\gamma, \; e^+, \; e^-$.
 A coincidence between the two tanks in the
 station is a shower signal.
\begin{figure}[thb]
\vspace*{-10pt}
\centerline{\includegraphics[width=70truemm]{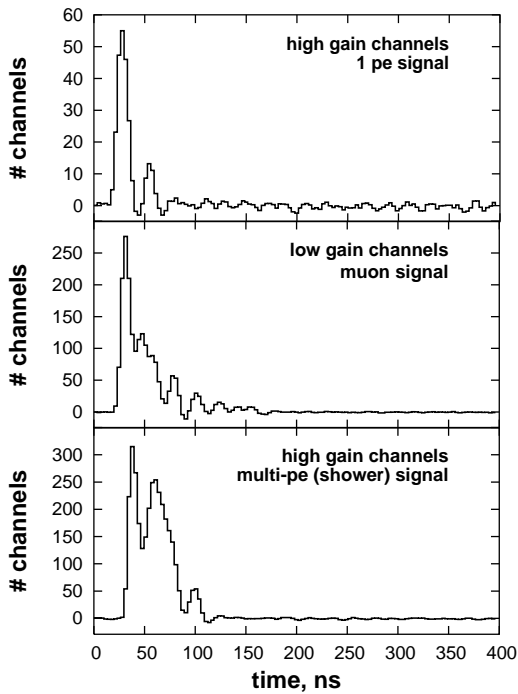}}
\vspace*{-20pt}
\caption{Samples of different types of waveforms, from top
to bottom: single \protect{\em pe}, most likely a muon, and  
 multi-\protect{\em pe},
 probably shower event.
\label{david}
}
\vspace*{-20pt}
\end{figure}
On board firmware recognizes 2-tank station coincidences above a
threshold of 30 MeV deposited energy. All such hits are transmitted to
the counting house, where a shower trigger is constructed from four-fold
station coincidences. This would correspond to a shower threshold energy
of 300 TeV for proton induced showers and about 500 TeV for iron induced
showers. We also keep isolated muons, recognized on board by a
combination of amplitude and shape. These muons are used for calibration
work and for identifying very large horizontal air showers by coincidence
across the array. Periodically the hit threshold will be lowered to take
single {\em pe} (photoelectrons) calibration data.

Figure~\ref{david} shows examples of waveforms extracted
 with early versions of the electronics from the test
tanks at South Pole for different types of events. The waveform from each
DOM is digitized at three different amplifications allowing for a dynamic
range of $1-10^3$~{\em pe} with $\sim$6 bits S/N. The sampling
time for the waveforms is set to 3.3 ns. Overall timing accuracy of a hit
is $< 10$ns including clock distribution to the DOMs.
The multi-{\em pe} \& muon waveform shapes are a convolution of single {\em pe}
waveforms with photon arrival times at the DOM. For the 2003/04 test
tanks the integrated signal from the low gain channels is 5-10\% that of the
high gain channels. The multi-{\em pe} waveform in the figure contains about 12
{\em pe}, and the muon $\sim 120$ {\em pe}.

The InIce and IceTop DAQs are being developed as an integrated IceCube
experiment. Either an IceTop or InIce trigger will cause a readout of the
entire experiment to build common events for offline analysis. This will
facilitate the collection of sub-threshold surface activity in
coincidence with in-ice muons, tagging a fraction of those muons as
atmospheric in origin. Such muons can be used to verify in-ice filter and
reconstruction algorithms.





\vspace*{-1truemm}
\section{SIMULATIONS}

 The IceTop simulation code is an integral part of the IceCube
 simulation. Parts of the code are already well developed.
 Specific items include\\
 Corsika shower simulations and transfer of the simulation
 data in a format convenient for both IceTop and InIce devices.
 This is achieved by an object oriented approach which allows 
 data exchange between the different
 simulation stages, similarly to the data exchange of real
 data~\cite{glacier}. 
 
 We have done the first run to calculate the event rates in
 the test tanks. Systematic cosmic ray shower simulations are
 to start in the beginning of 2005. \\

\begin{figure}[thb]
\centerline{\includegraphics[width=70truemm]{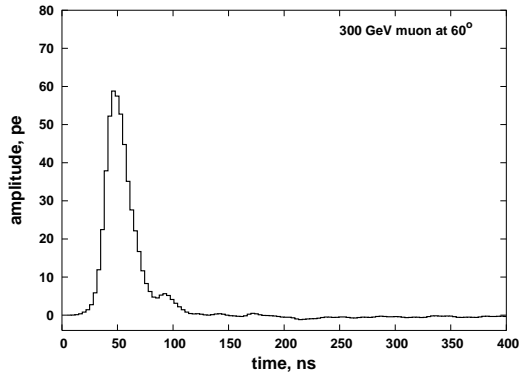}}
\vspace*{-20pt}
\caption{Simulated muon waveform of a 300 GeV muon that propagates
through the whole tank under zenith angle of 60\protect$^\circ$.
Note that the single \protect{\em pe} shape of the specific DOM is used to
create the waveform.
\label{wave}
}
\vspace*{-20pt}
\end{figure}

 We have constructed a tank simulation code
 based on GEANT 4~\cite{geant4}
 that generates the number of {\em pe} in the DOMs. This code is 
 complete and is currently used for verification by comparison 
 with test tank signals. We are using the results of the 
 GEANT 4 code to generate a set of photon tables for use
 in production simulations, suitable for
 treatment of the Corsika showers. 
 Simulation of the electronics is also in progress.
 We have also created an initial reconstruction code
 for treatment of the simulation results and for analysis
 of data from the first season. 

\section{DEPLOYMENT}

 The first IceCube deployment season is the austral summer of
2004/2005.  The plan is to deploy up to 4 strings and 
4 IceTop stations for a total of 8 tanks.  As of December 2004
all eight tanks are deployed, equipped with two DOMs each,
 filled with water and in the process of freezing. The tanks
will be closed for DOM operation in mid-January.
Freezing will be completed after the tanks are closed.
The entire 80-string, 160-tank IceCube detector is scheduled for installation
over the following five polar seasons.

Considering the first season as an engineering run, we expect to
verify the operation by measuring coincidence rates between
IceTop and deep detector.  Spacing between the two tanks at a 
station is optimized to select small showers by requiring
a coincidence between two tanks at the station with no hits
in adjacent stations.  Such showers correspond
to primary cosmic rays with energies 
of order 10-100 TeV which typically produce single muons
in the deep detector.  With 4 strings and 8 tanks we expect
a coincidence rate of approximately 0.3 Hz for such events.
With only four stations, this event sample will be contaminated
by large showers that fall outside the 4-station array.
In future years, as the array grows, these will be removed
and we expect to obtain a fairly clean sample of tagged
single muon events which will permit study of the main background
in the neutrino telescope.  The trajectory will be determined by
connecting the hit station with the center of gravity of the
deep detector signal.

The threshold for triggering the full 4-station air shower array is
approximately 300-500 TeV depending on the primary.  With a four-station
array we expect 4-station air shower triggers at a rate of
approximately 0.2 Hz.  The rate for 3-folds is approximately 0.3 Hz.
Taking into account the geometry factor for 4 strings instrumented
from 1.4 to 2.4 km and separated by 125 m, we expect a coincidence
rate between 4-fold air showers and muon bundles inside the 4-string IceCube of
approximately 20 per hour.

{\bf Acknowledgments} I appreciate the contributions of T.K.~Gaisser,
P.~Nie{\ss}en, D.~Seckel and S.~Tilav to this presentation.

\end{document}